\documentstyle[multicol,aps,prl,epsf]{revtex}

\begin{document}

\draft

\title{Hydrogen-assisted stabilization of Ni nanowires in solution}

\author{Manabu Kiguchi$^{1}$, Tatsuya Konishi$^{1}$, 
and Kei Murakoshi$^{1,2}$}

\address{$^1$Department of Chemistry, Graduate School of Science, 
Hokkaido University, Sapporo, 060-0810, Japan}
\address{$^2$PRESTO, Japan Science and Technology Agency, 
Sapporo, 060-0810, Japan}

\date{\today}

\maketitle

\begin{abstract}
We have studied conductance characteristics of mechanically fabricated Ni 
nanoconstrictions under controlling electrochemical potential and pH of the electrolyte. 
Conductance histogram showed clear feature peaked at 1-1.5 $G_{0}$ (=$2e^{2}/h$) when the potential 
of the constriction was kept at more negative potential than -900 mV vs. Ag/AgCl in pH=3.7. 
Comparable feature also appeared at more positive potential when lower pH solution was 
used. We have revealed that Ni mono atomic contact or mono atomic wire can be stabilized in 
solution at room temperature under the hydrogen evolution.\end{abstract}

\medskip

\begin{multicols}{2}
\narrowtext

Construction of metal nanowires is in the central issue of the current technology, because 
the techniques may be applied to fabricate ultimate electrical circuits in an atomic scale. 
Electrical conductance through a metal nanowire in an atomic scale is expressed by 
$G=2e^{2}/h \sum T_{i}$ ; where $T_{i}$ is the transmission probability of the $i$-th conductance channel, 
$e$ is the electron charge, and $h$ is Plank's constant. Conductance quantization depends not only 
on the atomic structure of the nanoconstriction but also on inherent properties of the 
metal \cite{1}. Conductance at mono-atomic point contact of transition metals with partial occupied $d$ 
orbitals is expected to be 1.5-3 $G_{0}$ based on the theoretical calculation \cite{1}. Recently, quantized 
conductance of Ni nanoconstrictions has attracted much attention because of the expectation 
that ferromagnetic nanowire may shows the conductance quantization in units of 0.5 $G_{0}$, since 
there is no spin degeneracy. The up- and down-spin electrons contribute independently to the 
electric transport. As well as Ni, other ferromagnetic metals, Fe and Co, have been 
extensively studied to confirm spin-dependent conductance quantization. However, 
previously documented data at experiments was less consistent each other. Various 
conductance values, such as 1.6 $G_{0}$ \cite{2}, 1 $G_{0}$ \cite{3}, and 0.5 $G_{0}$, 
have been reported \cite{4}. In addition, 
histogram of the conductance becomes featureless at room temperature \cite{5}. The poor agreement 
between former reported results may originate from the chemical instability of the 
nanoconstrictions of ferromagnetic metals. Novel experimental approach under controlling the 
chemical reactivity of ferromagnetic metals is required to study the characteristics of the 
conductance quantization of ferromagnetic metals. 

Electrochemical method has been recognized to be also another powerful approach to 
study the conductance quantization \cite{6,7}. Electrochemical potential determines the potential 
energy of electrons of the nanoconstriction, resulting in the control of the bonding strength 
between the metal atoms, and the interaction of the metals with molecules in surrounding 
medium. The characteristics of the system lead to successful fabrication of very stable metal 
nanostructures showing the conductance quantization which cannot be observed in UHV. For 
example, we reported that clear conductance quantization for Pd nanoconstrictions can be 
observed by the careful control of Pd deposition and dissolution\cite{6}. It should be noteworthy that 
conductance quantization was not observed in UHV at room temperature \cite{8}.
The experiments were performed in an electrochemical cell mounted in a chamber 
that was filled with high purity N$_{2}$ gas ($>$99.999$\%$) to avoid effect of oxygen in air. The tip 
was made of a gold wire (diameter $\sim$0.25 mm, $>$99.9$\%$) coated with wax to reduce Faradic 
current between the tip and a counter electrode. The substrate was Au(111) prepared by flame 
annealing and quenching method. The electrochemical potential ($\Phi_{0}$) of the Au substrate and 
tip was controlled using potentiostat (Pico-Stat, Molecular Imaging Co.) referring a Ag/AgCl 
reference electrode. The potential difference between the tip and substrate (bias voltage) was 
kept at 20 mV. A Pt wire was used as counter electrode. The electrolyte was 10 mM NiSO$_{4}$, 10 
mM H$_{3}$BO$_{3}$ and H$_{2}$SO$_{4}$. Solution pH was adjusted by changing 
the concentration of H$_{2}$SO$_{4}$. 

Nanoconstrictions of Ni were prepared as the following manner. First, Ni was 
electrochemically deposited both on the Au tip and the substrate by the polarization to 
negative potential than $\Phi_{0}$ = -650 mV where Ni bulk deposition proceeded \cite{9}. After the 
confirmation of the bulk deposition of Ni on the tip and a substrate, the tip was pressed into 
the substrate and then pulling out of contact at a typical rate of 50 nm/s in the electrochemical 
cell. During the contact breaking, a Ni nanoconstriction was formed between the tip and the 
substrate as schematically shown in Fig. 1. Conductance was measured by applying the bias 
of 20 mV between a tip and a substrate. 

Figure 2(a) shows the conductance histogram observed at $\Phi_{0}$ = -800 mV in the 
electrolyte solution with pH=3.7. There was no clear feature in the conductance histogram, 
suggesting that any certain atomic configurations were not preferentially formed between 
Ni-deposited tip and a substrate. When the electrochemical potential of the nanoconstriction 
was kept at more negative ($\Phi_{0}$ = -900 mV), clear feature appeared near 1-1.5 $G_{0}$ (see Fig. 2(b)). 
Further polarization to negative potential of the tip and the substrate did not change the 
conductance histograms. The clear feature in the conductance histogram showing 1-1.5 $G_{0}$ 
also appeared when the solution pH became lower than 2.7. Figure 2 (c) shows the 
conductance histogram at $\Phi_{0}$  =-800 mV in pH=2.3. Comparable feature as that of at $\Phi_{0}$= -800 
mV and pH=3.7 was clearly observed in the histogram. Further decrease in pH did not change 
the conductance histogram. The potential in which clear feature at 1-1.5 $G_{0}$ was observed 
corresponds to the hydrogen evolution region at Ni electrode \cite{10}. Thus, the present result shows 
that the chemical reaction of hydrogen evolution affects significantly the conductance of Ni 
nanoconstriction in solution. 

At the hydrogen evolution potential, the surface of Ni electrode is covered by the 
dissociated atomic hydrogen \cite{10}. The coverage of adsorbed hydrogen was evaluated by $in-situ$ 
infrared-reflection spectroscopy and the qualitative analysis of current-potential curves for 
hydrogen evolution \cite{10,11}. The coverage of adsorbed hydrogen gradually increases as the 
electrochemical potential becomes negative. Based on the results at the Ni polycrystalline 
electrode, the coverage of adsorbed hydrogen was estimated to be 0.20 and 0.40 at $\Phi_{0}$ = -800 
and -900 mV in pH=3.7, respectively. In more acidic solution with pH = 2.3, comparable 
coverage can be achieved at more positive potential by 100 mV. Comparable feature in the 
histograms in pH = 3.7 at -900 mV and that in pH = 2.3 at -800 mV proves that the 
dissociated atomic hydrogen play an important role to define the geometrical and/or the 
electronic structure of the ferromagnetic metal nanoconstriction in solution. 

It should be noteworthy that the observed conductance histogram (Fig. 2 (b)) was 
similar to that observed in UHV at ultra low temperature. In UHV, clear 1-1.5 $G_{0}$ feature 
appears in the conductance histogram at ultra low temperature \cite{12}, while the histogram becomes 
entirely featureless at room temperature \cite{5}. Thermal fluctuation of atoms at room temperature 
may prevent the long-term stability of the nanocontact, resulting in the broad feature in the 
histogram. At low temperature in UHV, thermal fluctuation is small. Thus the nanostructure 
can change into a thermally stable one showing specific values of the conductance. Although 
the fluctuation of atoms in solution at room temperature should be much larger than that in air 
and UHV, the specific feature of the Ni nanoconstriction was found to be kept during the 
hydrogen evolution reaction even at room temperature. The results prove that the control of 
electrochemical potential and solution pH leads to successful preparation of the Ni 
nanoconstrictions showing clear conductance quantization at room temperature.
 
At the present stage, it is not clear how the adsorbed atomic hydrogen stabilize the 
certain atomic arrangement of the Ni nanoconstriction. The result proves that the effects of 
hydrogen on the conductance values at the present system are different from those at UHV. In 
the previously documented result at UHV, introducing hydrogen suppressed the characteristic 
conductance of Ni at 1-1.5 $G_{0}$ by the replacement due to the evolution of a new value appears 
near 1 $G_{0}$ in the histogram \cite{12}. The origin of the value is a single hydrogen molecule bridging 
between two electrodes \cite{13}. In the case of Pd, introducing hydrogen also suppresses the 
characteristic conductance of 1.8 $G_{0}$, giving new values appear near 0.5 and 1 $G_{0}$ in the 
histogram \cite{14}. These values of 0.5 and 1 $G_{0}$ are attributable to a bridging single hydrogen 
molecule, and a Pd$_{2}$H$_{2}$ complex, respectively. On the other hand, at the present system in 
solution, the conductance histogram under the hydrogen evolution reaction was similar to that 
of a bare Ni metal in UHV. The observation suggests that the hydrogen molecules did not 
bridge between the electrodes, despite of the existence of the adsorbed hydrogen at the Ni 
surface. 

Electrochemical potential control of the Ni nanoconstriction in solution leads to the 
control of the electronic density of the Ni atoms at the constriction. Characteristics of the 
electronic structure of the Ni nanoconstriction under the electrochemical potential control 
should be maintained as the same as a bare Ni constriction in UHV, even at the situation in 
which hydrogen adsorbed on the surface. As well as the effect of potential control of the 
system, the adsorbed hydrogen is known to alter the surface stress, because of modulation of 
the bonding strength between Ni atoms at the surface \cite{15}. In the present system, adsorbed 
hydrogen as well as evolved hydrogen molecules in the vicinity of the Ni nanoconstriction 
may contribute to stabilize the specific structure of the constriction mechanically via 
relatively weak physical and/or chemical interactions. These specific effects in the 
electrochemical system may lead to comparable feature of the conductance histogram 
observed at the nanoconstriction of a bare Ni in UHV at ultra-low temperature, even under the 
hydrogen evolution in solution at room temperature. 

The conductance trace contains structural information on the Ni nanoconstriction in 
solution. Figure 3(a) shows the typical conductance trace at $\Phi_{0}$ = -800 mV in pH=3.7. As 
general tendency, the conductance change occurred continuously, rather than a stepwise 
fashion. Although conductance occasionally showed small steps and plateau-like structures, 
their appearance and conductance values were non-reproducible, resulting in the featureless 
conductance histograms (see Fig. 2 (a)). Figure 3 (b) and (c) show the typical conductance 
trace at $\Phi_{0}$ = -800 mV in pH=2.3. Although conductance change still occurred in a continuous 
fashion as at $\Phi_{0}$ = -800 mV in pH=3.7, the conductance of 1-1.5 $G_{0}$ was preferentially taken. It 
is the reason of an appearance of broad 1-1.5 $G_{0}$ feature in the histogram shown in Fig. 2 (c). 

At $\Phi_{0}$ = -800 mV in pH=2.3, the plateau was stretched as long as 0.8 nm (see Fig. 3 
(b)), and reversible transition between 1 $G_{0}$ and 0.6-0.7 $G_{0}$ were occasionally observed in the 
long plateau (see Fig. 3(c)). The conductance of a mono atomic contact is evaluated to be 
around 1.5-2.5 $G_{0}$ for transition metals \cite{1}. Therefore, the long plateau indicates high 
stabilization of the Ni mono atomic contact in solution. At the present stage, it is not clear 
whether the long plateau originated from the formation of the Ni mono atomic wire or it 
originated from deformation in a stem part of the nanoconstriction. This long plateau, 
however, implies the possibility of indicates the formation of a mono atomic Ni wire in 
solution. The formation of a mono atomic wire was supported by the reversible transition 
between 1 $G_{0}$ and 0.6-0.7 $G_{0}$ (Fig. 3(c)). A similar reversible transition of conductance 
between 1 $G_{0}$ and 0.6-0.7 $G_{0}$ were reported for Au mononanoatomic wires in the presence of 
physically adsorbed hydrogen in UHV at low temperature (10-30 K) \cite{16}. The reversible 
transition are explained by the dynamical structure transition between a dimerized Au wire 
and an equal-spacing wire, and the dimerization is a characteristic of a mono atomic wire \cite{17}. 

In conclusion, the dependence of the electrochemical potential and the solution pH 
on the conductance value was studied for the electrochemical deposited Ni nanoconstrictions. 
Under hydrogen evolution reaction, clear 1-1.5 $G_{0}$ feature appeared in the conductance 
histogram, suggesting that certain mono atomic arrangements were stabilized in solution. The 
analysis of the conductance trace showed the possibility of the formation of a 
one-dimensional Ni mono atomic wire in solution at room temperature. This 
hydrogen-assisted stabilization of metal nanostructure can be applied as an effective method 
to control the structure of the nanoconstriction in an atomic scale, and fabricate various 
interesting structures for the study of one-dimensional systems.

This work was partially supported by the Grant-in-Aid for Scientific Research from the 
Ministry of Education, Culture, Sports, Science and Technology, Japan.

\begin{figure}
\begin{center}
\leavevmode\epsfysize=55mm \epsfbox{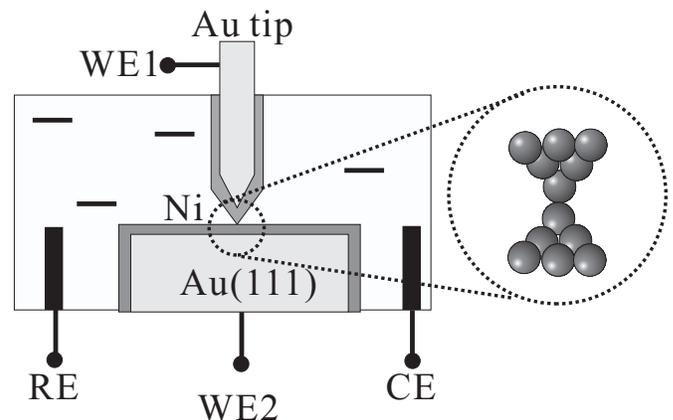}
\caption{Schematic presentation of a Ni nanoconstriction prepared under 
electrochemical-potential control. Ni is deposited on a Au single crystal substrate and a Au tip. 
RE, WE, and CE indicate reference electrode, working electrode, and counter electrode, 
respectively.}
\label{fig1}
\end{center}
\end{figure}

\begin{figure}
\begin{center}
\leavevmode\epsfysize=50mm \epsfbox{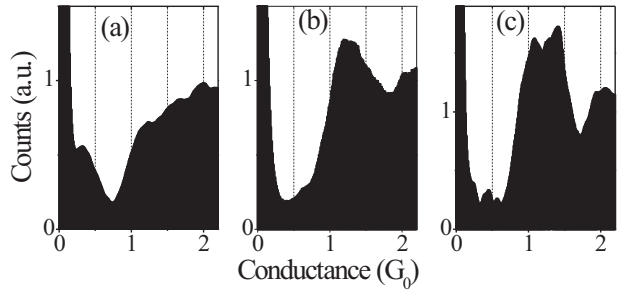}
\caption{Conductance histograms of Ni nanoconstrictions kept at (a, c) $\Phi_{0}$ = -800 mV, (b) 
$\Phi_{0}$ = -900 mV. Solution pH of the electrolytes was (a, b) 3.7 and (c) 2.3.}
\label{fig2}
\end{center}
\end{figure}

\begin{figure}
\begin{center}
\leavevmode\epsfysize=50mm \epsfbox{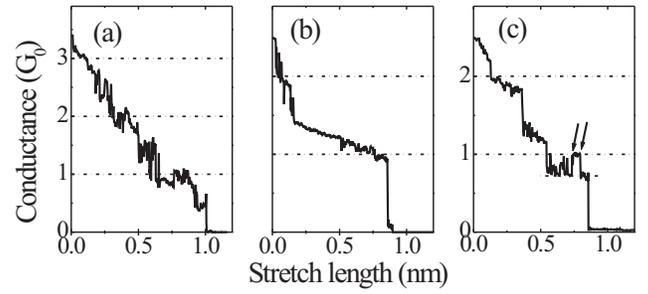}
\caption{Typical conductance traces of Ni nanoconstrictions observed at $\Phi_{0}$ = -800 mV in 
pH=3.7 (a), and pH=2.3 (b, c). Arrows in Figure 3 (c) indicate reversible transition between 1 
$G_{0}$ and 0.6-0.7 $G_{0}$ plateaus. }
\label{fig3}
\end{center}
\end{figure}

\end{multicols}
\end{document}